# Fluctuation-mediated spin-orbit torque enhancement in the noncollinear antiferromagnet Mn$_3$Ni$_{0.35}$Cu$_{0.65}$N

Arnab Bose[1*], Tom G. Saunderson[1,2], Aga Shahee[1], Lichuan Zhang[3], Tetsuya Hajiri[4], Adithya Rajan[1], Dongwook Go[1,2], Hidefumi Asano[4], Udo Schwingenschlögl[5], Aurelien Manchon[6], Yuriy Mokrousov[1,2], Mathias Kläui[1,7#]

[1]*Institute of Physics, Johannes Gutenberg-University Mainz, Staudingerweg 7, Mainz, 55128, Germany*
[2]*Peter Grünberg Institut and Institute for Advanced Simulation, Forschungszentrum Jülich and JARA, Jülich, 52425, Germany*
[3]*School of Physics and Electronic Engineering, Jiangsu University, Zhenjiang 212013, China*
[4]*Department of Materials Physics, Nagoya University, Nagoya, 464-8603, Japan*
[5]*Physical Science and Engineering Division, King Abdullah University of Science and Technology, Thuwal, 23955-6900, Saudi Arabia.*
[6]*Aix-Marseille Université, CNRS, CINaM, Marseille, France*
[7]*Centre for Quantum Spintronics, Norwegian University of Science and Technology, 7491 Trondheim, Norway*

\* *abose@uni-mainz.de. Current address: abose@iitk.ac.in*
\# *klaeui@uni-mainz.de*

**Abstract**

The role of spin fluctuations near magnetic phase transitions is crucial for generating various exotic phenomena, including anomalies in the extraordinary Hall effect, excess spin-current generation through the spin-Hall effect (SHE), and enhanced spin-pumping, amongst others. In this study, we experimentally investigate the temperature dependence of spin-orbit torques (SOTs) generated by Mn$_3$Ni$_{0.35}$Cu$_{0.65}$N (MNCN), a member of the noncollinear antiferromagnetic family that exhibits unconventional magnetotransport properties. Our work uncovers a strong and nontrivial temperature dependence of SOTs, peaking near the Néel temperature of MNCN, which cannot be explained by conventional intrinsic and extrinsic scattering mechanisms of the SHE. Notably, we measure a maximum SOT efficiency of 30%, which is substantially larger than that of commonly studied nonmagnetic materials such as Pt. Theoretical calculations confirm a negligible SHE and a strong orbital Hall effect that can explain the observed SOTs. We propose a previously unidentified mechanism wherein fluctuating antiferromagnetic moments trigger the generation of substantial orbital currents near the Néel temperature due to the emergence of scalar spin chirality. Our findings present an approach for enhancing SOTs, which holds promise for magnetic memory applications by leveraging antiferromagnetic spin fluctuations to amplify both orbital and spin currents.

**Introduction**

Spin-orbit torques (SOTs) currently provide a highly efficient mechanism for current-induced switching of magnetic elements as required for the implementation of next-generation spintronics devices such as nonvolatile magnetic random-access memories (MRAM)[1,2]. Typically, SOTs are generated by applying an in-plane electric current in the nonmagnet (NM)/ferromagnet (FM) bilayers, which generates transverse spin and orbital currents from the NM due to the spin-Hall effect (SHE)[3] and orbital-Hall effect (OHE)[4], respectively. Primarily the spin-orbit coupling (SOC) of NM has been exploited to produce an efficient transverse spin-Hall current ($J_{SH}$), that arises from the large intrinsic spin Berry curvature of the band structure[5–8] and extrinsic scattering mechanisms such as skew scattering and side jump[3,9–11]. Recently a nontrivial temperature dependence of $J_{SH}$ has been theoretically predicted for conducting ferromagnets originating in fluctuations of the magnetic moments[11], which was also experimentally reported[12–14]. The fluctuations of the spins have also been exploited to probe the magnetic transition of antiferromagnets via non-local magnon transport[15–17], spin-pumping[18], and the spin Seebeck effect[17,19]. These observations raise a fundamental question about whether one can utilize spin fluctuations of chiral antiferromagnets to directly produce large SOTs for practical applications. In this work, we report a huge SOT efficiency of $0.3 \pm 0.03$ generated by the noncollinear antiferromagnet (NC-AFM) $Mn_3Ni_{0.35}Cu_{0.65}N$ (MNCN) around its Néel temperature (210 K), which is much larger than that of conventional NM such as Pt[20]. Our observation of a large and nontrivial temperature dependence of the torques exerted by MNCN in the absence of any heavy element further suggests the possibility of a new physical mechanism, namely the fluctuations-mediated enhanced orbital-Hall current ($J_{OH}$).

Fluctuations of the magnetic order parameter near magnetic phase transitions are an important research area of spintronics, leading to various interesting effects such as a strong temperature dependence of anomalous Hall effect (AHE) and potentially an efficient approach to enhance the SHE. Additionally, spin fluctuations have been recently shown to have a drastic effect on the orbital properties of conducting electrons in magnets[21,22]. However, to unveil the fundamental mechanisms behind these effects, it is essential to investigate a system where magnetic layers are decoupled in order to discriminate the complex temperature-dependent contribution from the interfaces, which has rarely been reported[12–17]. At the same time, it is of fundamental interest to directly quantify SOTs in the absence of heavy elements to probe the role of spin fluctuations unambiguously while minimizing the contribution of the regular SHE. We overcome both these challenges by implementing MNCN/Cu/Py/Cu heterostructures. MNCN is a special type of NC-AFM that shows a sizable AHE driven by octupolar moments with vanishing magnetization[23–27] and produces various types of exotic transport properties[10,28–30]. A key advantage of MNCN is that its Néel temperature ($T_N$) is around 210 K, providing full access to quantify SOT in these heterostructures from cryogenic temperatures to room temperature.

**Experimental results**

We prepare high quality single crystal MNCN films of a thickness of 15 nm by the reactive magnetron sputtering technique on a (111) oriented single crystal MgO substrate (Fig. 1a). More details of the thin-film growth can be found in the methods section and previous works[26,31]. After the MNCN growth, Cu (1.5 nm) /Py (4-6 nm) / Cu (2 nm) / AlO$_x$ (2 nm) is deposited without breaking vacuum. We use a 1.5 nm Cu spacer between MNCN and Py to magnetically decouple these layers while the 2 nm thick top Cu layer is used to reduce the Oersted field generated by bottom layers for a cleaner detection of the signals in our experiment. In this material, the Kagome plane resides in the (111) sample plane where Hall-bars ($50 \times 20 \ \mu m^2$) are fabricated by the standard optical lithography, etching and lift-off techniques.

To quantify SOTs, we adopt the well-established technique of second harmonic Hall (SHH) measurements[32–36] performed in a three-dimensional-vector-cryo set up in presence of an external magnetic field ($H_{ext}$) at different temperatures ($T$) ranging from 4 K to 300 K. The schematics of the experiment is shown in Fig. 1b. We apply a low frequency (613 Hz) alternating electric current (ac) and record both the first harmonic ($V_{1\omega}$) and second harmonic Hall ($V_{2\omega}$) voltages while rotating $H_{ext}$ of different strength (0.03 T to 0.9) T, well above the in-plane saturation field of Py (<0.005 T). MNCN produces two different types of torques i.e. (1) In-plane damping-like torque (DLT), $\boldsymbol{\tau_{DL}} \propto \boldsymbol{m} \times (\boldsymbol{\sigma} \times \boldsymbol{m})$ where $\boldsymbol{m}$ is the unit vector of Py magnetization and $\boldsymbol{\sigma}$ is generated by angular momentum from SHE and/or OHE. For the DLT, the equivalent current-induced effective spin-orbit field (SOF), $H_{DL}^Z$ deflects the magnet out of the plane, creating an oscillation in the resistance due to AHE. (2) Out-of-plane field-like torque (FLT), $\boldsymbol{\tau_{FL}} \propto \boldsymbol{m} \times (H_{FL}^{Oe} + H_{FL}^Y)$ where $H_{Oe}$ is the oersted field generated by the conductive shunting layers and $H_{FL}^Y$ corresponds to interfacial SOF[36] which is found to be negligible and not the main focus of this work. FLT from $H_{FL}^{Oe}$ deflects the magnet in the sample plane resulting in oscillating resistance due to planar Hall effect (PHE). The mixing of the oscillating applied current and oscillating resistance produces $V_{2\omega}$, which can be expressed by equation (1)[32–36] in the absence of any other unconventional torques.

$$V_{2\omega} = C_A \cos\phi + C_P \cos\phi \cos 2\phi \tag{1}$$

Where
$$\begin{cases} C_P = -(H_{FL}^{Oe} + H_{FL}^Y)\frac{V_P}{H_{ext}} + C_0 \\ C_A = -H_{DL}^Z \frac{V_A}{2(H_{ext}+H_\perp)} + V_{ANE} + V_{ONE}H_{ext} \end{cases} \quad (2)$$

$V_{ANE}$ and $V_{ONE}$ are the strength of anomalous Nernst effect (ANE) and ordinary Nernst effect (ONE) respectively, which are generated due to the unintentional out-of-plane thermal gradient that is coupled to the in-plane magnetization of Py and $H_{ext}$ respectively[34]. $V_P$ is the coefficient of the PHE voltage, $V_{PHE} = V_P \sin 2\phi$ where $\phi$ is the angle between applied current and magnetization in the sample plane ($xy$ plane) (Fig. 1b). $V_P$ is determined by measuring $V_{1\omega}$ while rotating $H_{ext}$ in the plane (Fig. 1c). $V_A$ is the coefficient of the AHE voltage, $V_{AHE} = V_A \cos\theta$ where $\theta$ is the angle between the magnetization and out-of-plane, $z$-axis (Fig. 1d). The coefficient $V_A$ is obtained from $V_{1\omega}$ while sweeping $H_{ext}$ out of the plane (Fig. 1d). The same measurement also provides us with the information of $H_\perp$, the out-of-plane demagnetization field which is in the range of 0.4 T for a 4 nm thick Py film (Fig. 1d). More details can be found in the supplementary information.

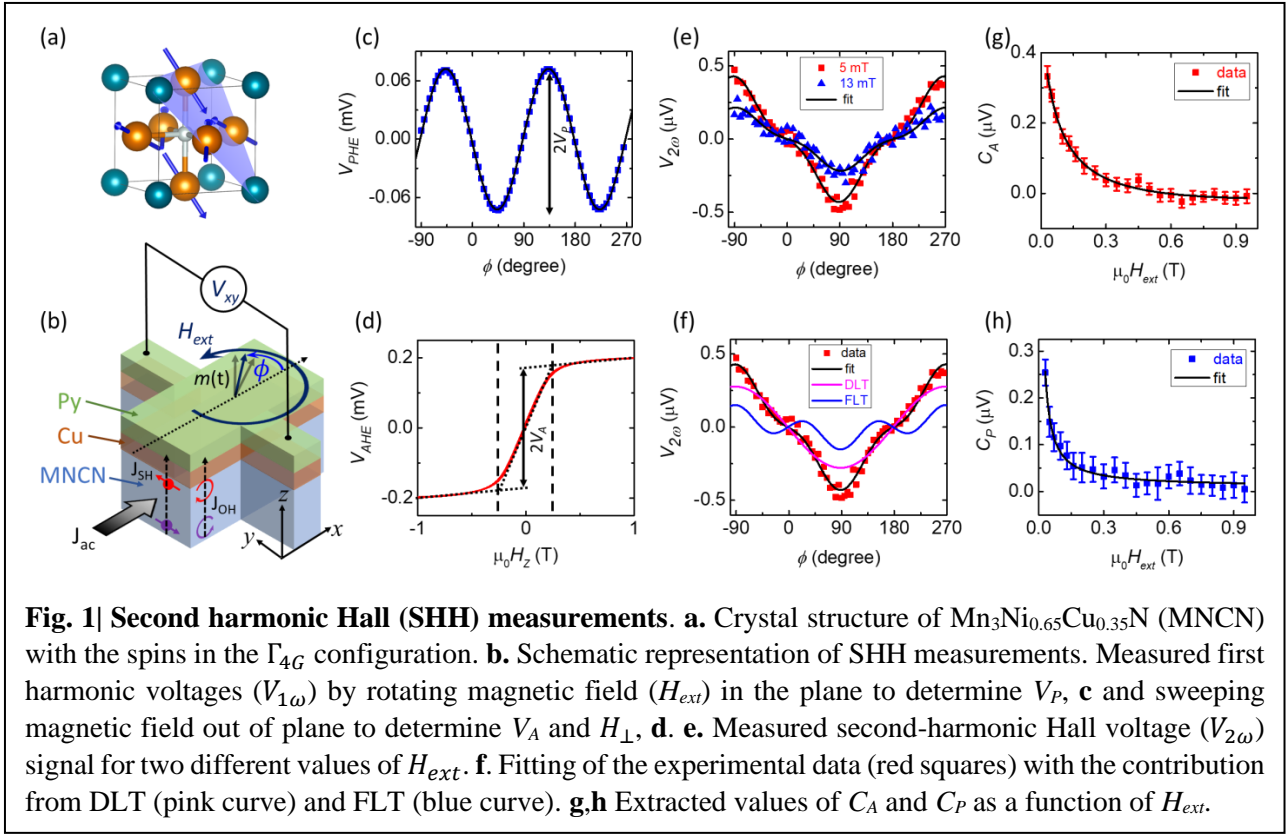

**Fig. 1| Second harmonic Hall (SHH) measurements**. **a.** Crystal structure of Mn$_3$Ni$_{0.65}$Cu$_{0.35}$N (MNCN) with the spins in the $\Gamma_{4G}$ configuration. **b.** Schematic representation of SHH measurements. Measured first harmonic voltages ($V_{1\omega}$) by rotating magnetic field ($H_{ext}$) in the plane to determine $V_P$, **c** and sweeping magnetic field out of plane to determine $V_A$ and $H_\perp$, **d**. **e.** Measured second-harmonic Hall voltage ($V_{2\omega}$) signal for two different values of $H_{ext}$. **f.** Fitting of the experimental data (red squares) with the contribution from DLT (pink curve) and FLT (blue curve). **g,h** Extracted values of $C_A$ and $C_P$ as a function of $H_{ext}$.

**Measurements of SOTs**

Fig. 1e shows a typical second harmonic voltage ($V_{2\omega}$) measured in our experiment for two different values of $H_{ext}$, 0.05 T (red squares) and 0.13 T (blue triangles), which are fit to equation (1) (black curves). We find that the magnitude of the measured $V_{2\omega}$ decreases for higher values of $H_{ext}$ suggesting that the signals predominantly originate from the SOTs (Eqn. 1-2). Unlike other similar types of single crystal AFMs, we do not observe any large contribution of unconventional torques[37,38] in our $\Gamma_{4g}$ configuration, which could be due to the presence of domain variants. Fig. 1f shows the fitting components of $V_{2\omega}$ for $H_{ext} = 0.05$ T using equation (1). The $\cos\phi$ component (pink curve) indicates a dominant contribution from DLT whereas $\cos\phi\cos2\phi$ component (blue curve) indicates a FLT. To quantify the current-induced spin-orbit fields, $H_{DL}^Z$ and $H_{FL}^Y$, we plot these fitting coefficients, $C_A$ (red squares) and $C_P$ (blue squares) as a function of $H_{ext}$ which fit well to equation (2) (black curve in Fig 1e,f). We find that both $C_A$ and $C_P$ decrease for higher magnitude of $H_{ext}$ and saturate close to zero suggesting SOTs are the source of the $V_{2\omega}$ signal and thermal signals, e.g. ANE and ONE are much weaker. Note that it is very important to perform the in-plane SHH measurement by rotating $H_{ext}$ up to a high value ($H_{ext} = 0.9\ T$), much larger than the demagnetization field ($H_\perp \sim 0.4\ T$ in our case) - to get an accurate estimation of $H_{DL}^Z$[34]. The DLT efficiency per unit electric field ($\xi_{DL}^E$) and per unit current density ($\xi_{DL}^j$) can be obtained from equation (3).

$$\begin{cases} \xi_{DL}^E = -\frac{2e}{\hbar}\mu_0 M_S t_{FM} \frac{H_{DL}^Z}{E} \\ \xi_{DL}^j = \xi_{DL}^E \rho \end{cases} \quad (3)$$

$\mu_0$ is the vacuum permeability, $\hbar$ is Planck's constant, $e$ is the electronic charge, $M_S$ is the saturation magnetization as obtained from the SQUID magnetometry (600 emu/cc ≈ 0.75 Tesla), $\rho$ is the electrical resistivity of MNCN (Fig. 3a). $\xi_{DL}^E$ and $\xi_{DL}^j$ are lower bounds of the internal spin-Hall conductivity ($\sigma_{SH}$) and spin-Hall angle ($\theta_{SH}$) respectively, due to the losses of the angular momentum transfer at the interface. To understand the origins of the SOTs we next probe its temperature dependence.

**Temperature dependence of SOT**

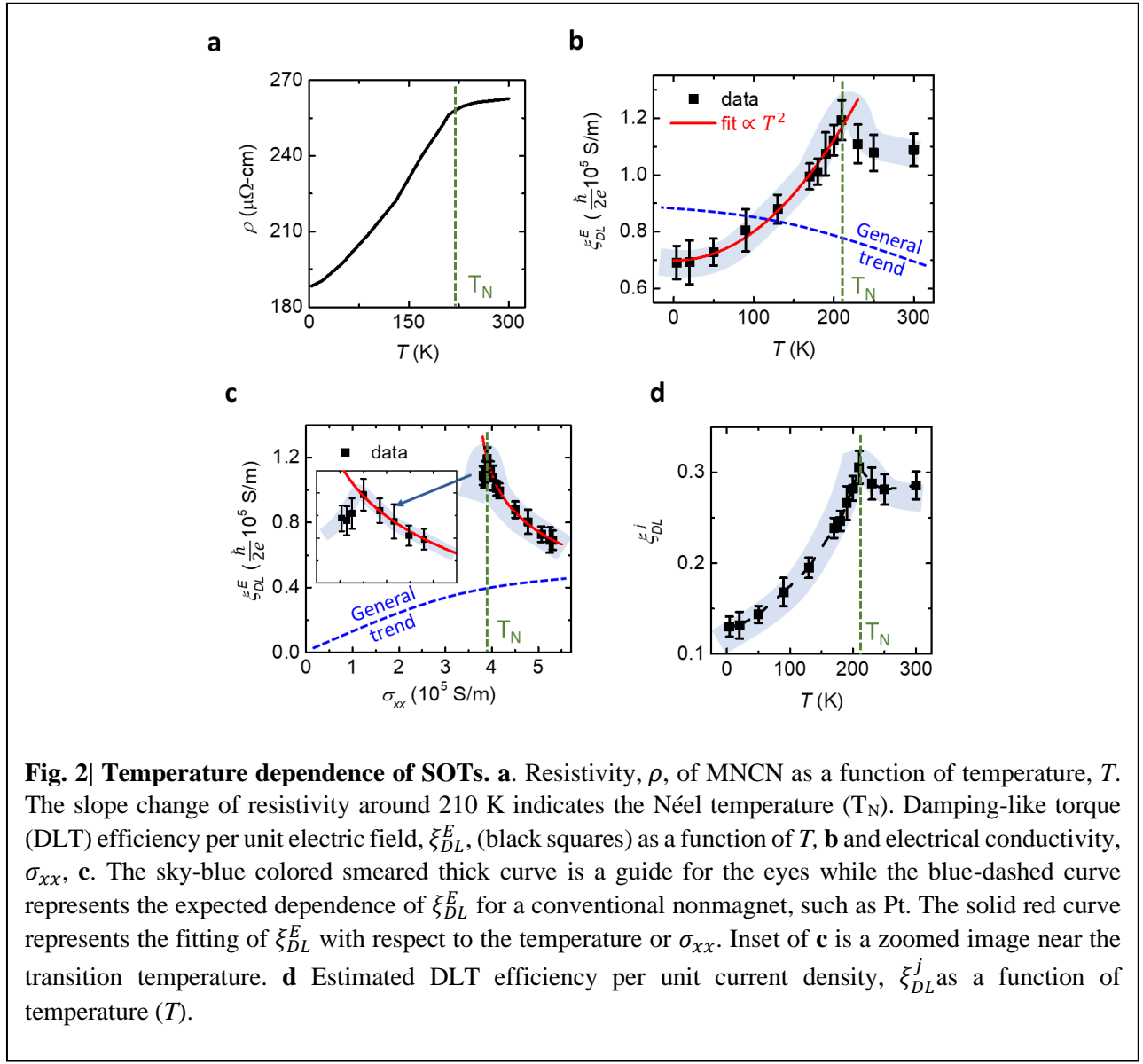

**Fig. 2| Temperature dependence of SOTs. a**. Resistivity, $\rho$, of MNCN as a function of temperature, $T$. The slope change of resistivity around 210 K indicates the Néel temperature ($T_N$). Damping-like torque (DLT) efficiency per unit electric field, $\xi_{DL}^E$, (black squares) as a function of $T$, **b** and electrical conductivity, $\sigma_{xx}$, **c**. The sky-blue colored smeared thick curve is a guide for the eyes while the blue-dashed curve represents the expected dependence of $\xi_{DL}^E$ for a conventional nonmagnet, such as Pt. The solid red curve represents the fitting of $\xi_{DL}^E$ with respect to the temperature or $\sigma_{xx}$. Inset of **c** is a zoomed image near the transition temperature. **d** Estimated DLT efficiency per unit current density, $\xi_{DL}^j$ as a function of temperature ($T$).

MNCN exhibits a metallic behavior showing an increase in the resistivity ($\rho$) with an increase of the temperature (Fig. 2a). There is a prominent change in $\rho$ around 210 K, suggesting a transition from the antiferromagnetic phase to paramagnetic phase which is also consistent with the literature[26,27,31]. Fig. 2b,c show a strong and unusual variation of $\xi_{DL}^E$ (black squares) with respect to both temperature ($T$) and longitudinal electrical conductivity ($\sigma_{xx}$) as highlighted with the sky-blue color thick curve. There is a large enhancement in both $\xi_{DL}^E$ (nearly 70 %) and $\xi_{DL}^j$ (approximately 140 %) near $T_N$ (Fig. 2b-d), strongly suggesting an influence of antiferromagnetic spin-fluctuations. The maximum ascertained values of $\xi_{DL}^E = \frac{e}{2\hbar}(1.2 \pm 0.1)10^5 S/m$ and $\xi_{DL}^j$=0.30 $\pm$ 0.03, which are much larger than the commonly studied heavy metals such as Pt[20] ($\xi_{DL}^j$=0.07) and, most surprisingly, in the absence of any heavy element in the MNCN compound. The general trend of $\xi_{DL}^E$ and $\sigma_{xx}$ as a function of temperature for a typical nonmagnet (such as Pt) is shown by the blue dashed line in Fig. 2b,c[3,39,40]. For the commonly used nonmagnets, we expect roughly $\sigma_{DL}^E \propto \sigma_{xx}$ in the "dirty metal regime" (for $\sigma_{xx} < 10^6 \, S/m$) due to the reduction of the carrier life time when the mean free path becomes comparable to the lattice constants[3,39,40]. As the temperature is lowered, when $\sigma_{xx}$ increases to the range of $10^6 - 10^8 \, S/m$ ("clean metal regime"), $\sigma_{DL}^E$ becomes nearly independent of $\sigma_{xx}$ as the carrier life time becomes comparable to the SOC energy[3,39,40]. For the ultra-clean metals ($\sigma_{xx} > 10^8 \, S/m$) when other spin-independent scattering processes are greatly supressed, we can observe the contribution from the extrinsic spin dependent scattering potentials producing $\sigma_{DL}^E \propto \sigma_{xx}^n$ where $n$>1[3,39]. Below the Néel temperature ($T_N$), in our case, $\sigma_{DL}^E$ decreases as $\sigma_{xx}$ increases with the decrease of temperature (Fig. 2b,c), which is at odd with any of these proposed mechanisms.

**Discussion**

To ascertain the origins of our observed large DLT-efficiency, we carried out density functional theory (DFT) calculations of MNCN in both nonmagnetic phase and NC-AFM phase with the static "all in/out" magnetic configuration (Fig. 3a,b), followed by atomistic spin dynamics simulations (Fig. 3c). The DFT calculation predicts that the intrinsic spin-Hall conductivity ($\sigma_{SH}$) in this material is small, of the order of $\frac{e}{\hbar}1 \times 10^4 S/m$, but the orbital-Hall conductivity ($\sigma_{OH}$) is larger by an order of magnitude, $\frac{e}{\hbar}5 \times 10^5 S/m$ in both nonmagnetic and NC-AFM phase (Fig. 3a,b). This result suggests that OHE fundamentally originates from the crystal structure rather than from the static magnetic ordering. Hence it could be possible that the orbital torques[4,41,42] from OHE play a crucial role in producing a large $\xi_{DL}^E$ in Ni-rich Ni$_{80}$Fe$_{20}$ (Py), which is further corroborated by the observed sign of the torques.

Next, we discuss various possible mechanisms for the nontrivial temperature dependence of the torques that shows an enhancement near $T_N$ and follows roughly $\xi_{DL}^E \propto T^2$ below $T_N$. It clearly signifies that we have to go beyond the conventional picture of scattering-free processes. Reference 11 predicts a nontrivial temperature dependence of $\xi_{DL}^E \propto T^{3.3}$ below $T_N$ for a conducting collinear FM where the coupling between the conduction electron and the dynamically fluctuating local magnetic moments produces SHE via scattering mechanisms. In addition, the chiral NC-AFMs have been predicted to generate an excess spin-current driven by the fluctuations of noncoplanar magnetic clusters with nonzero solid angles[10], which can also lead to such an unconventional temperature dependence of $\xi_{DL}^E$ in our system near $T_N$. On the other hand, it is noteworthy that the orbital magnetism in magnonic devices has been theoretically predicted[21] and also experimentally demonstrated[22].

We expanded upon the framework introduced in reference 21 for MNCN, observing that the scalar spin chirality ($S_1.(S_2 \times S_3)$) resulting from fluctuating spins ($S_i$) increases up to $T_N$, followed by a rapid decline to zero (depicted by the blue curve in Fig. 3c). A previous work has demonstrated that such an augmentation in the scalar spin chirality (SSC) can generate orbital angular momentum (Ref. 21), subsequently coming hand-in hand with chitality-induced orbital Hall current. The red curve in Fig. 3c illustrates the static mean-field OHE variation originating from the static orbitals across the magnetic phase. Its functional form lies in the assumption of the direct proportionality to the order parameter's temperature dependence (See section S4 and

Eqn s11 in supplemenray materials for more details). Note that SSC would produce $\sigma_{OH} = 0$ at 0 K and this curve has been shifted vertically for clarity in comparison. The interplay between these processes produces a peak magnitude in the orbital-Hall current, as elucidated by the black curve in Fig. 3(c) suggesting a qualitative agreement with our experiments (Fig. 3c). It is important to note that the predicted OHE from scalar spin chirality is expected to be proportional to $T^2$ (see section S4 in in supplemenray materials) for low temperature which agrees with our experimental results shown in Fig. 3b.

This study unveils a novel mechanism for generating a significant orbital Hall current driven by fluctuating spins, elucidating the role of scalar spin-chirality in noncollinear antiferromagnetic ordering. While the OHE emerges as one of the predominant mechanisms influencing experimentally measured SOTs, investigating the theoretical evolution of both OHE and SHE in these types of NC-AFMs remains an exciting avenue for future research. This exploration considers the strong correlation between fluctuating spin, orbit, and magnons.

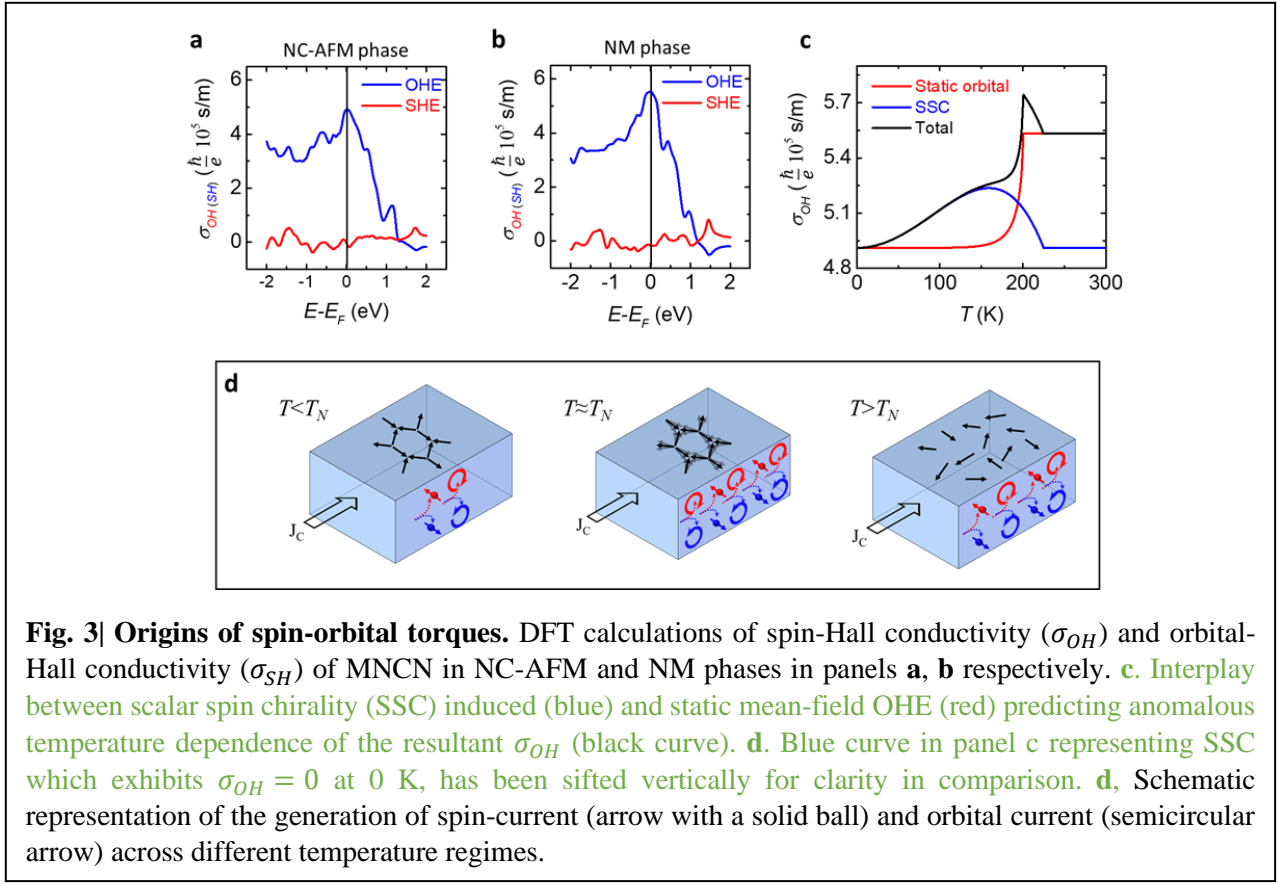

**Fig. 3| Origins of spin-orbital torques.** DFT calculations of spin-Hall conductivity ($\sigma_{OH}$) and orbital-Hall conductivity ($\sigma_{SH}$) of MNCN in NC-AFM and NM phases in panels **a**, **b** respectively. **c**. Interplay between scalar spin chirality (SSC) induced (blue) and static mean-field OHE (red) predicting anomalous temperature dependence of the resultant $\sigma_{OH}$ (black curve). **d**. Blue curve in panel c representing SSC which exhibits $\sigma_{OH} = 0$ at 0 K, has been sifted vertically for clarity in comparison. **d**, Schematic representation of the generation of spin-current (arrow with a solid ball) and orbital current (semicircular arrow) across different temperature regimes.

## Conclusion

We report an unprecedented temperature dependence of SOTs in the single crystal noncollinear antiferromagnet Mn$_3$Ni$_{0.35}$Cu$_{0.65}$N. There is a 140% enhancement in the DLT efficiency per unit current density ($\xi_{DL}^j$) around the Néel temperature that arises from the fluctuations of the antiferromagnetic order and that boosts the generation of both spin and orbital currents. The estimated $\xi_{DL}^j$ at the Néel temperature is approximately 0.3, significantly larger than that can be realized using conventional heavy metals such as Pt. This is remarkable considering that MNCN does not contain any heavy metal. Our theoretical calculations suggest that the observed effect could be attributed to the fluctuation-enhanced OHE rather than SHE. The potential generation of large orbital-Hall currents exploiting fluctuating spin textures can open up a new avenue for the MRAM application by achieving large SOTs.

**Acknowledgments**

A.B. is thankful to Alexander von Humboldt foundation for the postdoctoral fellowship. M.K., A.S., and A.B. thank the Graduate School of Excellence Materials Science in Mainz (MAINZ, GSC266); Spin+X (A01, A11, B02) TRR 173-268565370 and Project No. 358671374; the Horizon 2020 Framework Programme of the European Commission under FETOpen Grant Agreement No. 863155 (s-Nebula); the European Research Council Grant Agreement No. 856538 (3D MAGiC); and the Research Council of Norway through its Centers of Excellence funding scheme, Project No. 262633 "QuSpin.". A.R, and M.K acknowledge funding from the European Union's Framework Programme for Research and Innovation Horizon 2020 (2014 − 2020) under the Marie Sklodowska-Curie Grant Agreement No. 860060 (ITN MagnEFi).T.G.S., D.G., and Y.M. gratefully acknowledge the Jülich Supercomputing Centre for providing computational resources under project jiff40. L.Z thank to the funding from the National Natural Science Foundation of China (12347156), the Natural Science Foundation of Jiangsu Province (BK20230516). T.H and H.A acknowledge funding from the Japan Society for the Promotion of Science (KAKENHI Grant Nos. 20H02602 and 19K15445). U.S., A.M. and M.K, thank the funding from King Abdullah University of Science and Technology (KAUST) under award 2020 − CRG8 − 4048. A.M was supported by the Excellence Initiative of Aix-Marseille University A*MIdex, a French" Investissements d'Avenir program". In addition, the Deutsche Forschungsgemeinschaft (DFG, German Research Foundation) − Grant No. TRR 173/2 – 268565370

**Authors contribution**

T.H. grew the single crystal thin films under supervision of H.A. A.B. played the primary role in performing the device fabrication, measurement and data analysis under the supervision of M.K. A.S. and A.R assisted in the experiments. T.G.S. carried the theoretical calculations with inputs from D.G. and Y.M. The manuscript was written by A.B with the inputs of M.K, A.M, Y.M, and D.G. All the authors commented on the manuscript and contributed to this work. M.K. has been the principal investigator of this project.

**Methods**

**Sample preparation.** We prepare high quality MNCN films of thickness 15 nm by the reactive magnetron sputtering technique on a (111) oriented single crystal MgO substrate at temperature of 375° C in presence of Ar (96%) and $N_2$ (4%) gases, with partial pressure of 2 Pa. After the growth, the thin films are annealed at 500° C to obtain the single crystal structure (Fig. 1a). More details of the thin-film growth can be found in previous work[26,31]. Following the cooling to room temperature, Cu (1.5 nm) /Py (4-6 nm) / Cu (2 nm) / $AlO_x$ (2 nm) is successively deposited without breaking vacuum. We use a 1.5 nm Cu spacer between MNCN and Py to magnetically decouple these layers, while the 2 nm thick top Cu layer is used to reduce the Oersted field generated by bottom layers for a cleaner detection of the signals in our experiment. The top $AlO_x$ is employed for capping to prevent the heterostructure from oxidation. In this material, the Kagome plane resides in the (111) sample plane where Hall-bars ($50 \times 20\ \mu m^2$) are fabricated by the standard optical lithography, etching and lift-off techniques.

**Transport measurement and data processing.** The second harmonic Hall measurements were performed in a 3d vector cryo setup using a lock-in amplifier (signal recovery 7265). A low-frequency current of 613 Hz was sourced to the device while an in-plane external magnetic field ($H_{ext}$) is rorated in the sample plane. Both first harmonic ($V_{1\omega}$) and second harmonic Hall voltages ($V_{2\omega}$) were measured. By fitting $V_{2\omega}$ to equation (1) and analyzing the field dependence of the coefficient of $\cos\phi$ term of equation (1), we quantify the DLT efficiency per unit electric field ($\xi_{DL}^E$). We repeat all these measurements for different values of the temperature and the final analysis is presented in Fig. 3. By multiplying the resistivity to $\xi_{DL}^E$, we obtain the DLT efficiency per unit current density ($\xi_{DL}^j$) which is the lower bound of spin-Hall angle of the material.

**Computational details.** The Density Functional Theory (DFT) calculations were performed for the bulk $Mn_3Ni_{0.35}Cu_{0.65}N$ considering the lattice constant 3.9012°[26] using Wannier90 package[43,44] interfaced with the FLEUR codes[45]. It implements the full potential linear augmented plane wave method (FP-LAPW)[46]. For the exchange correlation functional we chose the Perdew-Burke-Ernzerhof functional within the generalized gradient approximation[47]. More details are provided in the supplementary information.